\documentclass[preprint,12pt,3p,times]{elsarticle}

\usepackage{amssymb}
\usepackage{amsmath,amsfonts,amsthm,bm} 
\usepackage{multirow}
\usepackage{booktabs}
\usepackage{adjustbox}

\usepackage{multirow} 
\usepackage{multicol}
\usepackage{algorithm}
\usepackage{siunitx}
\usepackage{algpseudocode}
\usepackage{adjustbox}
\usepackage{algorithm}
\usepackage{algpseudocode}
\usepackage{xcolor}

\journal{Information Sciences}

\begin{document}

\begin{frontmatter}

\title{DecKG: Decentralized Collaborative Learning with \\Knowledge Graph Enhancement for POI Recommendation}

\author[label1,label2]{Ruiqi Zheng}\ead{12150084@mail.sustech.edu.cn}
\affiliation[label1]{organization={Department of Computer Science and Engineering, Southern University of Science and Technology},
            addressline={}, 
            city={Shenzhen},
            postcode={518055},
            country={China}}

\author[label1,label2]{Liang Qu}\ead{qul@mail.sustech.edu.cn}
\author[label3]{Guanhua Ye\corref{cor1}}\ead{g.ye@bupt.edu.cn}

\author[label2]{Tong Chen}\ead{tong.chen@uq.edu.au}

\affiliation[label2]{organization={School of Electrical Engineering and Computer Science, The University of Queensland},
            addressline={}, 
            city={Brisbane},
            postcode={4072}, 
            country={Australia}}

\affiliation[label3]{organization={School of Computer Science, Beijing University of Posts and Telecommunications},
            addressline={}, 
            city={Beijing},
            postcode={100876}, 
            country={China}}

\author[label1]{Yuhui Shi\corref{cor1}}\ead{shiyh@sustech.edu.cn}

\author[label2]{Hongzhi Yin\corref{cor1}}\ead{db.hongzhi@gmail.com}

\cortext[cor1]{Corresponding authors}

\begin{abstract}

Decentralized collaborative learning for Point-of-Interest (POI) recommendation has gained research interest due to its advantages in privacy preservation and efficiency, as it keeps data locally and leverages collaborative learning among clients to train models in a decentralized manner. However, since local data is often limited and insufficient for training accurate models, a common solution is integrating external knowledge as auxiliary information to enhance model performance. Nevertheless, this solution poses challenges for decentralized collaborative learning. Due to private nature of local data, identifying relevant auxiliary information specific to each user is non-trivial. Furthermore, resource-constrained local devices struggle to accommodate all auxiliary information, which places heavy burden on local storage. To fill the gap, we propose a novel decentralized collaborative learning with knowledge graph enhancement framework for POI recommendation (DecKG). Instead of directly uploading interacted items, users generate desensitized check-in data by uploading general categories of interacted items and sampling similar items from same category. The server then pretrains KG without sensitive user-item interactions and deploys relevant partitioned sub-KGs to individual users. Entities are further refined on the device, allowing client-client communication to exchange knowledge learned from local data and sub-KGs. Evaluations across two real-world datasets demonstrate DecKG’s effectiveness recommendation performance.
\end{abstract}




\end{frontmatter}



\section{Introduction}
\label{sec:intro}

Point-of-Interest (POI) recommendation systems help users discover locations of interest, such as restaurants or landmarks, based on their preferences and location history \cite{islam2022survey}. These systems are commonly used to improve location-based services, including mobile advertisements \cite{2021Discovering}. Typically, a centralized cloud server manages user data, as well as the training and inference of recommendation models \cite{2014GeoMF, zheng2023automl}. However, this raises privacy concerns, as users' sensitive interacted data are transferred to the cloud server, potentially violating privacy regulations such as the General Data Protection Regulation (GDPR)\footnote{https://gdpr-info.eu/}. Furthermore, heavy dependency on server infrastructure and steady internet connectivity \cite{2020Next} may impact the trustworthiness of cloud-based models.

This has led to the development of models that can be deployed directly on edge devices such as smartphones \cite{qi2022correlation, qi2018two, yin2024device}. This allows for local generation of recommendations with minimal reliance on centralized resources. Research in this field can be categorized into two main types: 
(1) Federated POI recommendations (FedPOIs) \cite{yan2024federated, yao2023fedrkg}, which primarily rely on server-client collaboration. A central server collects locally trained models, aggregates them with aggregation algorithms (e.g., FedAvg \cite{pmlr-v54-mcmahan17a}), and redistributes them to all clients. However, FedPOIs impose significant scalability bottlenecks, as the system's performance is constrained by the communication load and the computational burden placed on the central server. 
(2) Decentralized collaborative learning POI recommendations (DecPOIs) \cite{chen2018privacy,long2023model,long2023decentralized} which offer an alternative to FedPOIs by eliminating the heavy reliance on a central server. In this paradigm, clients collaborate directly with each other, sharing locally trained models with neighboring clients, rather than uploading models to a central server. The main responsibility of the central server is to identify communication neighbors by analyzing similarities in user preferences. This architecture is particularly suitable for POI recommendations, as it reduces the communication overhead and delays associated with central server optimization and redistribution, while also better capturing the interactions and influences among nearby users.

Despite the benefits of DecPOIs, the limited interaction history preserved on the device hinders performance. A common solution is integrating knowledge graph (KG) as auxiliary information to enhance model performance \cite{liu2023self,xuan2023knowledge}. For instance,  we could represent the relationship between an "Apple Official Store" and a "Retailer Store" with the triples (Apple Official Store, sells, Apple Products) and (Retailer Store, sells, Apple Products), respectively. Users who visit one of these stores are more likely to visit the other due to the shared entity (i.e., "Apple Products"), capturing the semantic relationships encoded within the auxiliary information. Although there are efforts to incorporate KG into FedPOIs, these approaches either deploy the entire KG on the device \cite{yao2023fedrkg}, overlooking their capacity limitations, or only focus on FedPOIs \cite{ma2024fedkgrec, yan2024federated}, lacking effectiveness in the context of DecPOIs. 
Since DecPOIs are particularly well-suited for POI recommendations by direct client-client communication, exploring how to effectively integrate KG into DecPOIs is crucial. However, with DecPOIs emerging as new methods, the challenges of effectively utilizing KG in this setting remain unexplored.

Integrating KG into DecPOIs presents several challenges due to the privacy-preserving need and  the constraints of local devices. Firstly, the server must determine user preferences to effectively assign communication neighbors, which raises concerns about how to upload user preferences without compromising sensitive information. Secondly, 
due to resource-constrains and computational limitations of the devices, it is impractical for server to deploy the entire KG on the device side. Identifying relevant sub-knowledge graph sub-KGs for each user without accessing the sensitive user check-in data is essential. Finally, current DecPOIs typically focus on sharing model parameters between neighboring clients, while the integration of sub-KGs into this client-client communication remains unresolved. Developing an effective method to share and incorporate sub-KGs across devices could improve recommendation performance by enriching local models with items' relational knowledge, without exposing sensitive user information.

To fill this gap, we propose a new framework: decentralized collaborative learning with knowledge graph enhancement framework for POI recommendation (DecKG). This framework leverages a server to identify user neighbor sets, facilitating collaborative learning among neighbors. The cloud server pretrains the entire knowledge graph, which contains items and other entities, excluding the user-item interactions. Each client employs a check-in data perturbation mechanism to generate desensitized check-in data by uploading the general category of the interacted item and sampling similar items from the same category, excluding the actual ones. Upon receiving the desensitized data from clients, the cloud server partitions the whole KG using meta-paths. Each entity in the desensitized check-in data is treated as a head entity to identify related tail entities within the KG. Once a tail entity is found, it is then treated as a new head entity to search for additional tail entities, continuing this process until no further entities can be identified. This iterative approach generates multiple meta-paths, which collectively form a user-specific sub-KG that encapsulates all relevant entities and relations. Then the server deploys only the sub-KGs relevant to each specific user. On the device side, all entities, including individual user embeddings, undergo further learning based on local training data and sub-KGs. Following local training, client-client communication promotes knowledge exchanges between neighbors. The contributions of this paper are summarized as follows:
\begin{itemize}
    \item We propose a decentralized learning paradigm for the knowledge graph. It pretrains without private user-item interactions and further refines through client-client communication between neighbors, without exploding sensitive user check-in data.
    \item We introduce the Decentralized Collaborative Learning with Knowledge Graph Enhancement (DecKG) framework for recommendation task. To the best of our knowledge, this is the first study to explore the integration of auxiliary information into DecPOIs. This approach not only enriches local models with external knowledge but also enhances the system's applicability in real-world scenarios, particularly in settings that prioritize privacy and decentralized operations.

    \item We execute comprehensive experiments of DecKG with two datasets collected from real-world scenarios, highlighting its effectiveness in enhancing recommendation performance. 

\end{itemize}



\section{Related Work}
This section reviews contemporary publications on Point-of-Interes (POI) recommendation, decentralized collaborative recommender systems, and recommendation with knowledge graph enhancement.
\label{sec:related work}

\subsection{Point-of-Interes (POI) Recommendation}
Early approaches to assist users in discovering appealing locations by interpreting user click-in histories primarily concentrated on Markov chains \cite{2013Where} and matrix factorization techniques \cite{2014GeoMF}. More recently,  recurrent neural networks (RNNs) have been used to learn POI check-in data \cite{2017SERM, 2018DeepMove, QiangLiu2016PredictingTN, 2019Where}. SGRec \cite{2021Discovering} builds on these methods by incorporating graph-enhanced POI sequences, effectively leveraging collaborative signals from semantically linked POIs while learning sequential patterns, yielding higher accuracy than RNN-based models. Additionally, attention-based models \cite{ 10.1145/3477495.3531983} employ self-attention mechanisms to capture relative spatiotemporal dependencies among check-in events. Attempts have also been made to utilize side information. RankGeoFM \cite{li2015rank} uses dual embeddings to represent both user preferences and location information, while MTEPR \cite{chen2021multi} integrates social networks and POI locations to enhance recommendation performance. Despite these advancements, these models operate in a centralized manner, collecting sensitive user data on the server side and potentially violating privacy regulations.

\subsection{Decentralized Collaborative Recommender Systems}

In contrast to federated recommender systems \cite{qu2024towards,yuan2024hetefedrec} that rely on a server for constant aggregation and distribution of public models, decentralized collaborative recommender systems \cite{chen2018privacy,long2023model,long2023decentralized} adopt a more efficient approach by limiting the server's role to the initial phase. At the initial stage, the server's primary task is to assign users to neighbors with similar preferences, followed by local optimization and client-client communication. DMF \cite{chen2018privacy} pioneered the use of Matrix Factorization in a decentralized framework, where neighbors are selected according to geographical proximity. Expanding on this, DPMF \cite{yang2022dpmf} employs a probabilistic MF model that incorporates both implicit and explicit user feedback to capture preferences and item features. For POI recommendations, DCLR \cite{long2023decentralized} enhances the neighbor selection process by including both semantic and geographic proximity, enabling users to collaborate with others likely to visit the same locations. MAC \cite{long2023model} introduces a user-centric neighborhood identification mechanism, allowing users not only to be assigned neighbors by the server but also to independently select high-quality neighbors for knowledge exchange, thereby further improving recommendation performance.

\subsection{Recommendation with Knowledge Graph Enhancement}
The recommendation with knowledge graph has gained traction due to its ability to model complex relational structures between users, items, and auxiliary entities. CKE \cite{zhang2016collaborative} enhances the collaborative filtering framework by integrating items' textual features and attributes using TransE \cite{bordes2013translating}, while CFKG employs TransR \cite{lin2015learning} for similar purposes. SHINE \cite{wang2018shine} approaches recommendation as a link prediction task within the knowledge graph, identifying potential connections between entities. Additionally, several approaches have refined user representations by incorporating both their interaction history and the multi-hop neighbors of the interacted items, allowing for the capture of high-order relational information. RippleNet \cite{wang2018ripplenet} pioneered the propagation of user preferences across knowledge graphs, and AKUPM \cite{tang2019akupm} extends this idea by leveraging TransR for entity representation and employing self-attention to assign propagation weights. Despite their success, all methods require access to sensitive user embeddings, making them infeasible to integrate into existing privacy-preserving recommendation frameworks.

\section{Preliminaries}
In this section, we present commonly used definitions in Point-of-Interest (POI) recommendation and knowledge graphs, as well as the problem formulation for the proposed method.

\textbf{Definition 1: Check-in Data.} For user $u_i \in \mathcal{U}$, $\mathcal{X}(u_i) = \{p_1, p_2, \dots, p_j\}$, consists of $n_j$ POIs visited by the user. Each POI $p_j \in \mathcal{P}$ is associated with a category $c_{p_j}$, consisting of the category data $\mathcal{X}^{c}(u_j) = \{c_{p_1}, c_{p_2}, \dots, c_{p_j}\}$, and the set of all categories is represented as $\mathcal{C}$. A segment $s$ represents a geospatial division of POIs, and the set of all segments is represented as $\mathcal{S}$.

\textbf{Definition 2: Knowledge Graph}. Given a knowledge graph $\mathcal{G} = \{ (h, r, t) \ | h,t \in \mathcal{E}; r \in \mathcal{R} \}$, where $\mathcal{E}$ is the set of entities, and $\mathcal{R}$ is the set of relations. Every triplet $(h,r,t)$ represents a relation $r$ between the head entity $h$ and the tail entity $t$. The set of entities $\mathcal{E}$ contains the POI $p_j$, POI category $c_{p_j}$, the segment $s_j$, and other side information such as the brand of the POI $p_j$. The set of Relations $\mathcal{R}$ depends on the specific dataset, such as whether the POI belongs to a brand or the regions are nearby to each other.

\textbf{Problem Formulation: Decentralized POI Recommendation with Knowledge Graph Enhancement.} For DecKG framework, the different roles of users/clients and cloud server are defined as follows:

\begin{itemize}
\item \textbf{User/Client}: For one user $u_i$ maintains distinct sensitive check-in data $\mathcal{X}(u_i)$, category data $\mathcal{X}^c(u_i)$, a sub-knowledge graph $KG_i$ assigned by the server, and a customized model $\phi_i(\cdot)$ trained collaboratively with local sensitive check-in data and refined through client-client communication. 

\item \textbf{Server}: The server's primary responsibility is to identify neighbor sets and partition the sub-knowledge graph $KG_i$ for each user $u_i$ in the initial stage. After distributing to users, the server stays inactive during the following local model training and communication process.
\end{itemize}

Leveraging local user-item interactions and the knowledge graph, the local model $\phi_i (\cdot)$ is learned in a privacy-preserving manner to predict the likelihood of user $u_i$ visiting POI $p_j$, denoted as $\hat{y} _{u_i,p_j}$.

\begin{figure*}[hbt!]
\centering
\includegraphics[width=1\textwidth]{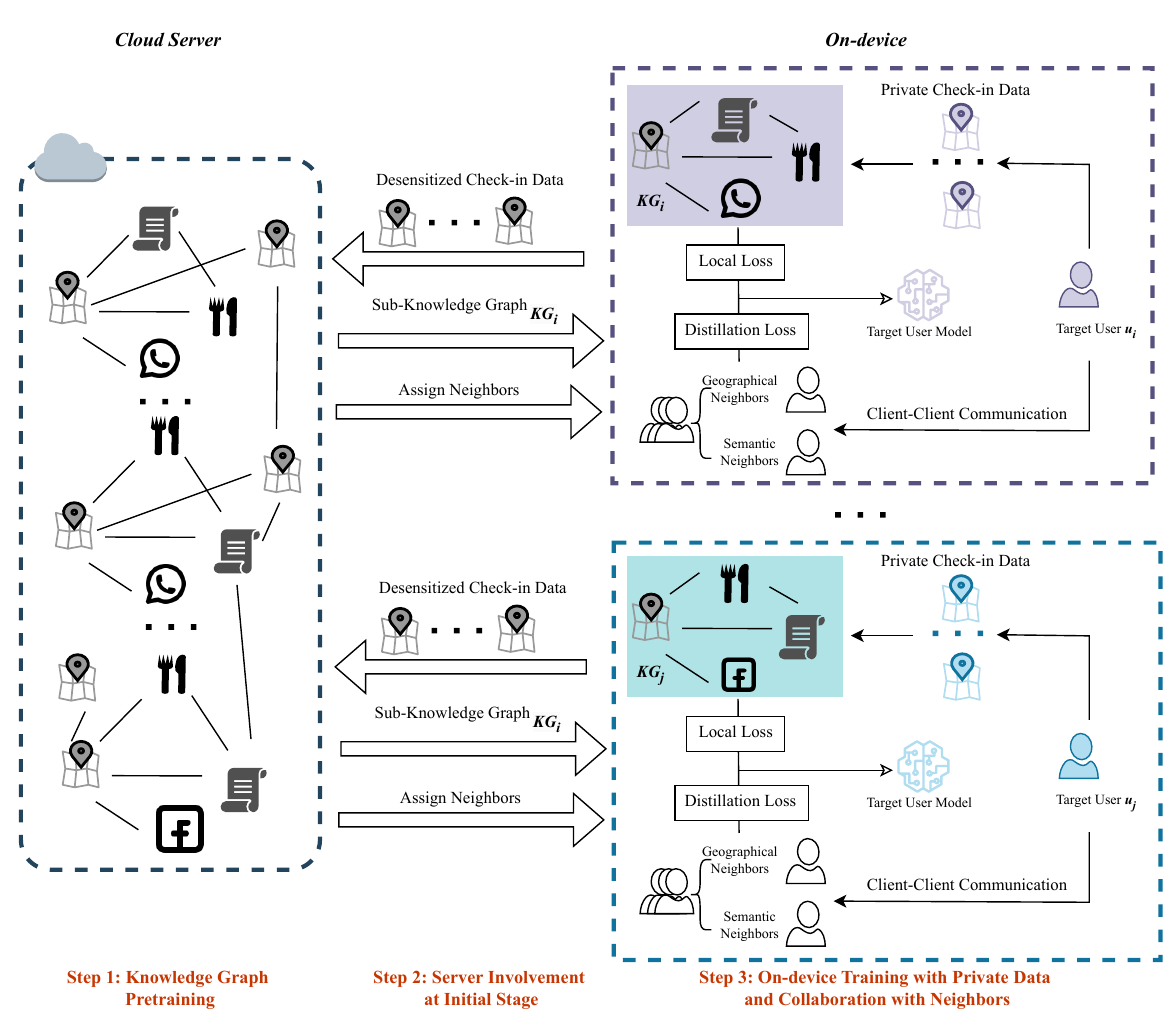} 

\caption{The overview of DecKG. a) Step 1: Knowledge graph is pretrained on the server side without the sensitive user-item interaction data. b) Step 2: At the initial stage, client upload the desensitized check-in data to the server, while server assign neighbors and partition the KG into $KG_i$ for user $u_i$. c) Step 3: Users are trained with the private check-in data combining the sub-KGs, and the knowledge from neighbors in client-client communication. }
\label{fig:overview}

\end{figure*}
\section{DecKG}

In this section, we present the framework of our proposed DecKG method, as depicted in Fig. \ref{fig:overview}. We begin by introducing the general DecPOI framework in Section \ref{sec:DecPOI}. Step 1, covering knowledge graph pretraining, is detailed in Section \ref{sec:pre}. In Step 2, we explain how users generate desensitized data in Section \ref{sec:generation}, followed by the server's role in partitioning the knowledge graph and assigning neighbors in Section \ref{sec:assign}. Finally, Step 3, which involves on-device refinement with client-client communication, is discussed in Section \ref{sec:refine}.

\subsection{Decentralized Collaborative Learning POI Recommendations}
\label{sec:DecPOI}

Following the widely adopted DecPOIs framework \cite{kermarrec2010RW,defiebre2020decentralized,defiebre2022human}, users maintain control over their sensitive data on local devices, where they train local user-specific models. The server is responsible to assign communication neighbors to each client, enabling decentralized collaborative learning through client-client knowledge sharing (e.g., model gradients).

In DecPOIs, each user $u_i$ manages a private local dataset $\mathcal{O}_i = (u_{i}, v_{ij}, r_{ij})_{j=\{1,...,|\mathcal{O}_i|\}}$, where $v_{ij}$ is the item interacted with by user $u_i$, and $r_{ij}$ is the rating. Every client retains a local model $\phi_i(\cdot)$, including item embeddings $\textbf{V}_i \in \mathbb{R}^{K \times |\mathcal{V}|}$, user embedding $\textbf{u}_i \in \mathbb{R}^{K}$, and other parameters $\Theta_i$. $K$ is the embedding dimension. Due to the limited local data, it is insufficient to train robust models. DecPOIs utilize a central server to assign neighbors to each user by calculating similarities between users using non-sensitive data, like local item embedding tables \cite{defiebre2022human,defiebre2020decentralized}. Through this setup, clients share their knowledge (e.g., item embedding gradients $\frac{\partial \mathcal{L}_{u_i}}{\partial \textbf{V}_i}$) with their assigned neighbors $\mathcal{N}(u_i)$ to collaboratively refine local model $\phi_i(\cdot)$, such as item embeddings $\textbf{V}_i$, through client-client communication as follows:
\begin{equation}
    \textbf{V}_i = \textbf{V}_i - \gamma Agg(\nabla\textbf{V}_k)_{u_{k} \in \mathcal{N}(u_i)}
\label{equ:CL}
\end{equation}

$\nabla\textbf{V}_k = \frac{\partial \mathcal{L}_{u_k}}{\partial \textbf{V}_k}$ and $\gamma$ is the predefined learning rate. $\mathcal{L}_{u_i}$ refers to the client’s loss function, and $Agg(\cdot)$ represents an aggregation function.

\subsection{Knowledge Graph Pretraining}
\label{sec:pre}
Instead of directly learning the KG on the device side, we first pretrain the full embedding of all entities on the server with a classical knowledge-based recommendation method and then refine it on the device side. It is worth noticing that our contributions focus on the DecKG learning paradigm, integrating the KG into the existing DecRecs frameworks. This pertaining process can be accomplished by any knowledge-based model, from embedding-based methods \cite{lin2015learning, bordes2013translating} to path-based methods \cite{wang2019explainable, yu2014personalized}. we
employ a general graph neural network model (i.e., KGCN \cite{wang2019knowledge}) as the pertaining based model since it captures multi-hop high-order
information and shows competitive performance under various scenarios. 
For a triplet $(h,r,t)$ and a negative sampling triplet $(h,r,t^{'})$, the pairwise ranking loss is determined as: 

\begin{equation}
\mathcal{L}_{KG} = \sum_{(h,r,t) \in \mathcal{G}, (h,r,t^{'}) \notin \mathcal{G}} - \ln \sigma \left(score(h,r,t^{'}) -  score(h,r,t)\right), 
\end{equation}
where $\sigma(\cdot)$ is the activation function and $score(h,r,t) = || \textbf{e}_r + \textbf{W}_{r}^{\mathsf{T}} \textbf{e}_{h} - \textbf{W}_{r}^{\mathsf{T}} \textbf{e}_t ||^{2}_{2}
$ represents the distance between the projection of the tail entity embedding $\textbf{e}_t \in \mathbb{R}^{d}$ and the summation of the head entity embedding $\textbf{e}_h \in \mathbb{R}^{d}$ and the relation embedding $\textbf{e}_r \in \mathbb{R}^{k}$. $\textbf{W}_r \in \mathbb{R}^{d  \times k}$ is the transformation matrix, which maps the entities to the relation space. For an entity $t$, the output of 
$l$-th layer is the aggregation of its previous layer embedding and the embedding of its neighbors:

\begin{equation}
\begin{aligned}
\textbf{e}^{l}_{t} &= \sigma \left( \textbf{W}^{\mathsf{T}} (\textbf{e}^{l-1}_t + \sum_{h \in \mathcal{G}_{t}} \eta_{th} \textbf{e}_h^{l-1} ) + \textbf{b}\right) \\
\eta_{th} &= (\sqrt{|\mathcal{G}_{t}| \cdot |\mathcal{G}_{h}|} \ \  )^{-1}
\end{aligned}
\end{equation}
where $\textbf{W}$ is the learnable transformation weight and $\textbf{b}$ is the bias. After the pertaining step, the whole KG is ready for further partition on the cloud server based on the received desensitized check-in data.

\subsection{Desensitized Check-in Data Generation}
\label{sec:generation}
To facilitate the server in assigning neighbors and partitioning knowledge graphs for users, it is essential that users disclose their preferences without revealing sensitive interactions. One straightforward approach is to apply the Random Response \cite{dwork2014algorithmic} technique to perturb each user-item interaction list. Specifically, for each interaction $(u_i, p_j)$ where $p_j \in \mathcal{P}$, the user reverses whether they clicked on the item with probability $p = \frac{1}{1 + e^{\epsilon}}$. However, this method may affect the learning performance of the knowledge graph \cite{qin2017generating} and since the user only clicks on a small portion of the POI set, this method makes the knowledge graph denser, introducing excessive noise. Therefore, we generate desensitized check-in data by selecting the item \( p^{'}_j \) from a candidate set consisting of POIs of the same category, \( \mathcal{P}^{'}(u_i) = \{p_m \mid c_{p_m} = c_{p_j}; p_m \neq p_j\} \), to replace the target POI \( p_j \) in the user interaction history \( \mathcal{X}(u_i) \). It is important to note that replacing items across different categories could drastically affect performance, and thus, the substitution is constrained to items within the same category. The selection process leverages the similarity between the embeddings of the candidate POI and the target POI, utilizing a probability distribution to balance relevance and diversity in the recommendations. The similarity between the embedding of a candidate item \( \mathbf{e}_{p_m} \) and that of the target item \( \mathbf{e}_{p_j} \) is measured using cosine similarity, defined as follows:

\begin{equation}
\begin{aligned}
    sim(\mathbf{e}_{p_m}, \mathbf{e}_{p_j}) &= \frac{\mathbf{e}_{p_m} \cdot \mathbf{e}_{p_j}}{\|\mathbf{e}_{p_m}\| \|\mathbf{e}_{p_j}\|}, \\
    Pr(p_m) &= \frac{\exp(\epsilon \cdot sim(\mathbf{e}_{p_m}, \mathbf{e}_{p_j}))}{\sum_{{p_k} \in \mathcal{P}^{'}(u_i)} \exp(\epsilon \cdot sim(\mathbf{e}_{p_k}, \mathbf{e}_{p_j}))}.
\end{aligned}
\label{equ:similarity_and_probability}
\end{equation}

To determine the probability \( Pr(p_m) \) of selecting a candidate item \( p_m \) as a replacement for the target item \( p_j \), we adopt an exponential mechanism, which assigns higher probabilities to items with greater similarity while ensuring a level of randomness that encourages diversity in the selection process. The parameter \( \epsilon \) controls the sensitivity of the selection mechanism. A higher \( \epsilon \) value makes the selection more deterministic, favoring items with greater similarity, whereas a lower \( \epsilon \) value introduces more randomness. The process is repeated $|\mathcal{X}(u_i)|$ times to generate the desensitized check-in data $\mathcal{X}^{'}(u_i)$.

\subsection{Knowledge Graph Partition and Neighbor Identification}
\label{sec:assign}

For each user $u_i$, the desensitized check-in data $\mathcal{X}^{'}(u_i)$ is uploaded to the server, which then generates a sub-knowledge graph tailored to the user for further refinement. A straightforward approach is to include the tail entities directly associated with the entities in the check-in data as one-hop neighbors, formulated as $KG_i = \{(h,r,t) \mid h \in \mathcal{X}^{'}(u_i); t \in \mathcal{E}; r\in \mathcal{R}\}$. However, this method does not capture multi-hop, high-order information. To address this, we form a meta-path. Each entity in the desensitized check-in data is treated as a head entity to find related tail entities within the KG. Once a tail entity is identified, it is treated as a new head entity, continuing the search for additional tail entities. This process iterates until no further entities can be identified.

This iterative approach generates multiple meta-paths, collectively forming a user-specific sub-KG that encapsulates all relevant entities and relations. Given $p_j \in \mathcal{X}^{'}(u_i)$, we construct the meta-path $Path(p_j) = p_j \xrightarrow{r_1} entity_2 \xrightarrow{r_2} \cdots \xrightarrow{r_n} entity_{n+1}$ to traverse all possible entities connected to the check-in $p_j$. The set $Entity(Path(p_j))$ contains all entities in the meta-path except the last one. The sub-KG is formulated as:

\begin{equation}
KG_i = \left\{ (h,r,t) \bigm| h \in \{Entity(Path(p_j)) \mid p_j \in \mathcal{X}^{'}(u_i)\}; t \in \mathcal{E}; r\in \mathcal{R} \right\}
\end{equation}

This sub-KG is constructed from all possible meta-paths generated by the desensitized check-in data and will be sent to the user’s device for further refinement.

Additionally, the server assigns both geographical neighbors and semantic neighbors to each user. Due to the locality of POI recommendation, users who frequently visit venues within the same geographical segment $s \in \mathcal{S}$ have high geographical affinity. Such users' information can be mutually beneficial for predicting future movements. We define these users as geographical neighbors of $u_i$, denoted by $\mathcal{N}_{geo}(u_i)$. For two users $u_i$ and $u_j$, if $u_j$ has visited $u_i$'s geographical segment, then $u_j$ is considered a geographical neighbor of $u_i$.

Semantic neighbors $\mathcal{N}_{sem}(u_i)$, on the other hand, refer to users who may be geographically distant but share similar preferences. The similarity between users is calculated based on their desensitized check-in data:

\begin{equation}
similarity(\mathcal{X}^{'}(u_i), \mathcal{X}^{'}(u_j)) = \left\| \frac{\sum_{p_m \in \mathcal{X}^{'}(u_i)} \mathbf{e}_{p_m}} {|\mathcal{X}^{'}(u_i)|} - \frac{\sum_{p_n \in \mathcal{X}^{'}(u_j)} \mathbf{e}_{p_n}} {|\mathcal{X}^{'}(u_j)|} \right\|^2_2
\end{equation}

For each user $u_i$, users with the high similarity score $similarity(\mathcal{X}^{'}(u_i), \mathcal{X}^{'}(u_j))$ are selected as semantic neighbors, and the number of semantic neighbors is equal to the corresponding user' geographical neighbors. The neighbor set for user $u_i$ is the combination of above two kinds of neighbors $\mathcal{N}(u_i) = \mathcal{N}_{sem}(u_i) \cup \mathcal{N}_{geo}(u_i)$.

\subsection{On-device Refinement}
\label{sec:refine}
After transferring the sub-knowledge graph and completing neighbor identification, the server remains inactive. On the device side, all entities, including individual user embeddings, undergo further learning based on local training data and sub-KGs. Following local training, client-client communication facilitates knowledge exchanges between neighbors.

\subsubsection{Local Loss Function}
Given that $\mathcal{X}(u_i)$ represents the sensitive POIs interacted histories, we use the local loss function to train and update the model parameters:
\begin{equation}
  \mathcal{L}_{loc}(u_i) = l\left(\phi_i\left(\mathcal{X}(u_i)\right), \mathcal{Y}(u_i)\right),
\end{equation}
where $l$ is the loss function, $\phi_i(\cdot)$ denotes the model for user $u_i$, and $\mathcal{Y}(u_i)$ is the ground truth. 

\subsubsection{Distillation Loss Function}
Both semantic and geographical neighbors have varying levels of influence, as their similarity to the local target user differs. An affinity-based model parameter aggregation method is utilized to transfer knowledges from neighbors through client-client communication. The aggregation weight is based on how similar the neighbor is compared to the target user:

\begin{equation}
\mathcal{L}_{nbr} = \sum_{u_j \in \mathcal{N}(u_i)}  \frac{similarity(\mathcal{X}^{'}(u_i), \mathcal{X}^{'}(u_j))}{\sum_{u_k} similarity(\mathcal{X}^{'}(u_i), \mathcal{X}^{'}(u_k))} \mathcal{L}_{loc}(u_j)
\end{equation}

The final loss function is a combination of the local loss and the distillation loss from neighbors. Both the user embeddings and entities in the sub-KGs are updated through this loss.$\mu$ is a hyperparameter that governs the balance between the aggregated model and current model:

\begin{equation}
\mathcal{L}_{total}(u_i) = (1-\mu) \mathcal{L}_{loc}(u_i) + \mu \mathcal{L}_{nbr}(u_i)
\end{equation}

\section{Experiments}
\label{sec:experiment}
To evaluate the effectiveness of DecKG, we conduct comprehensive experiments designed to address several key research questions (RQs):

\textbf{RQ1}: How does DecKG perform in comparison to federated and decentralized POI recommendation methods? This question examines the competitive advantages of DecKG over existing approaches, particularly in terms of recommendation accuracy and efficiency.

\textbf{RQ2}: How do different hyper-parameters influence the performance of DecKG? We explore the sensitivity of DecKG's performance to various hyper-parameter settings, aiming to understand how tuning specific parameters impacts the overall system.

\textbf{RQ3}: How do individual components of DecKG contribute to its overall performance? By isolating and analyzing the different elements of DecKG, we assess the significance of each in enhancing recommendation quality.

\textbf{RQ4}: Can DecKG be seamlessly integrated with other decentralized POI recommendation methods? This research question investigates the interoperability of DecKG with existing decentralized systems, evaluating its adaptability and potential for broader applications.


\subsection{Experimental Settings}

\subsubsection{Datasets}
Two frequently used real-world datasets to assess our proposed method DecKG: Beijing and Shanghai \cite{liu2021improving,hong2023urban}. Both datasets encompass users’ check-in histories (user-POI interaction) and corresponding knowledge graphs. Following \cite{liu2021improving} 10,000 users in each dataset are selected and data is split into 8:1:1 for training, testing, and validation. Table \ref{tab:dataset} presents the principal statistics of these datasets.

\begin{table}[htbp]
 \renewcommand{\arraystretch}{1}
  \centering
  
  \caption{An examination of datasets' statistical properties.}
\begin{adjustbox}{max width=0.6\textwidth}
\begin{tabular}{c|c|c|c|c|c|c}
\toprule
      & \#Users & \#POIs & \#Check-in & \#Entities & \#Relations & \#Triplets \\
\midrule
\midrule
Beijing & 10, 000  & 177, 602 & 763, 073 & 181, 817  & 20    & 336,178 \\
Shanghai & 10, 000  & 169, 123  & 618, 197  & 173, 504  & 20    & 384,702 \\
\bottomrule
\end{tabular}%
\end{adjustbox}
  \label{tab:dataset}%
\end{table}%

\subsubsection{Evaluation Protocols}
Two ranking metrics are used: Normalized Discounted Cumulative Gain at Rank $k$ (NDCG@$k$) and the recall at Rank $k$ (Recall@$k$).
NDCG@$k$ emphasizes the high ranks of the ground truth items that appear in the top-$k$ list, and Recall@$k$ evaluates the proportion of the ground truth items.

\subsubsection{Baselines}
Since POI recommendation is highly sensitive to user privacy, we will compare our proposed method with other privacy-preserving methods that retain user click histories on the client side rather than uploading them to the server. There are two primary categories of such methods: federated POI recommendations (FedPOIs) and decentralized collaborative POI recommendations (DecPOIs). We will select representative methods from each category, along with methods that integrate knowledge graphs with these approaches.

For FedPOIs, FedNCF \cite{perifanis2022federated} focuses solely on user-item interactions, while FedCFKG and FedCKE incorporate knowledge graphs. The specific details of these methods are described below:

\begin{itemize}
\item \textbf{FedNCF}: It is a federated recommendation framework of NCF \cite{he2017neural}. Clients use multilayer perception to learn the user-item interaction. User embedding is preserved on the client side, and item embedding is uploaded to the server for global synchronization. 

\item \textbf{FedCFKG}: It is a federated recommendation framework implementing CFKG \cite{ai2018learning} to retrieve and learn the knowledge graph.  TransE \cite{bordes2013translating} is utilized to learn the embedding of the entities. We implemented and modified ourselves to better suit the application and privacy-preserving protocol.

\item \textbf{FedCKE}: It is a federated recommendation framework implementing CKE \cite{zhang2016collaborative} to predict the existence of the triplets.  TransR \cite{lin2015learning} learns the embedding of the entities. Similarly, it has been modified to meet the privacy-preserving protocol. 

\end{itemize}

For DecPOIs, DCLR \cite{long2023decentralized} and DARD \cite{zheng2024decentralized} utilize interaction data, while the proposed method DecKG incorporates knowledge graphs into client-client collaborations. To our best knowledge, this is the first attempt to leverage the power of knowledge graphs for DecRecs. The specific details of these methods are described below.

\begin{itemize}
\item \textbf{DCLR}: It is a decentralized framework of on-device POI recommendation. Initially, the server will group users based on regions. Then clients are trained with local data and collectively learn from their geographic neighbors and semantic neighbors through exchanging model parameters. 

\item \textbf{DARD}: It is also a decentralized framework. It implements an adaptive mechanism to construct reference data as the bridge for client-client communication. Instead of directly exchanging raw parameters like DCLR, clients share their preference on the reference data to implicitly transfer learned knowledge.

\end{itemize}

\subsection{Top-$k$ Recommendation (RQ1)}

\begin{table}[htbp]
 \renewcommand{\arraystretch}{1}
  \centering
  
  \caption{The top-$k$ recommendation results.}
\begin{adjustbox}{max width=0.7\textwidth}

\begin{tabular}{c|c|c|c|c|c|c|c}
\toprule
\multirow{2}[4]{*}{Datasets} & \multirow{2}[4]{*}{Metric} & \multicolumn{3}{c|}{Federated Recommendation} & \multicolumn{3}{c}{Decentralized Recommnedation} \\
\cmidrule{3-8}      &       & FedNCF & FedCFKG & FedCKE & MAC   & DARD  & DecKG \\
\midrule
\midrule
\multirow{4}[8]{*}{Beijing} & NDCG@10 & 0.0273  & 0.0201  & 0.0375  & 0.0292  & 0.0314  & \textbf{0.0389 } \\
\cmidrule{2-2}      & Recall@10 & 0.0258  & 0.0178  & 0.0347  & 0.0225  & 0.0281  & \textbf{0.0354 } \\
\cmidrule{2-2}      & NDCG@20 & 0.0325  & 0.0240  & 0.0465  & 0.0339  & 0.0378  & \textbf{0.0474 } \\
\cmidrule{2-2}      & Recall@20 & 0.0372  & 0.0271  & 0.0526  & 0.0368  & 0.0435 & \textbf{0.0559 } \\
\midrule
\multirow{4}[8]{*}{Shanghai} & NDCG@10 & 0.0261  & 0.0176  & 0.0234  & 0.0256  & 0.0245  & \textbf{0.0321 } \\
\cmidrule{2-2}      & Recall@10 & 0.0252  & 0.0175  & 0.0215  & 0.0267  & 0.0271  & \textbf{0.0348 } \\
\cmidrule{2-2}      & NDCG@20 & 0.0336  & 0.0212  & 0.0278  & 0.0353  & 0.0342  & \textbf{0.0399 } \\
\cmidrule{2-2}      & Recall@20 & 0.0443  & 0.0253  & 0.0332  & 0.0459  & 0.0464  & \textbf{0.0512 } \\
\bottomrule
\end{tabular}%

\end{adjustbox}
  \label{tab:rq1}%
\end{table}%

To evaluate the performance of the proposed method DecKG, we compare it with carefully selected baselines for a fair comparison on the Top-$k$ recommendation task. The performance results are presented in Table \ref{tab:rq1}, and key observations are listed below.

The proposed method DecKG significantly outperforms all baseline methods on both datasets. The rich information embedded within the knowledge graph combined with the flexible decentralized collaborative learning framework enables DecKG to learn superior representations of users and items for the recommendation task. Due to the privacy-preserving protocols employed in DecKG, users retain their sensitive click histories on their devices, preventing unauthorized access. However, in the POI task, the items that users have visited are often limited and sparse. Both collaborative learning and knowledge graphs can mitigate the data sparsity problem through more facts about the items and others' perspectives on the items. The superior performance of DecKG compared to other models indicates that the data sparsity problem indeed poses a significant challenge for other approaches.

While knowledge graphs offer significant potential, they are not a panacea. The proposed method successfully integrated the knowledge graph into DecRecs framework, resulting in a notable performance improvement. However, even incorporating a knowledge graph, the performance of FedCFKG is not satisfactory compared with FedNCF. A simple inclusion of a knowledge graph does not guarantee a good performance. Employing appropriate learning strategies within a suitable framework is also crucial. The proposed method organically integrates the knowledge graph specifically for DecRecs through a two-step process: knowledge graph pretraining on the server side followed by refinement on each device tailored to client-client communication.

\subsection{Hyperparameter Study (RQ2)}

To investigate the effect of two crucial hyperparameters, we evaluate the performance for different values of $\epsilon = \{ 0.5, 2, 4, 8\}$ and $\mu = \{0.1, 0.3, 0.5, 0.7\}$, as shown in Figure \ref{fig:hyper}.

\textbf{Impact of $\epsilon$}. $\epsilon$ controls the sensitivity of the selection mechanism. A higher $\epsilon$ value makes the selection more deterministic, favoring items with greater similarity, whereas a lower $\epsilon$ introduces more randomness. When $\epsilon$ reaches 8, the performance improvement becomes less noticeable. This is likely because the mechanism has already selected sufficiently similar items to replace the sensitive ones, preserving the characteristics of the original check-in data while adhering to the privacy-preserving protocol.

\textbf{Impact of $\mu$}. $\mu$ controls the balance between the current model and the aggregated model based on the neighbors' information. Both extremely small and large $\mu$ values lead to performance degradation. This indicates that a proper balance between learning from local data and transferring knowledge from neighbors is essential for achieving optimal performance. Relying solely on local data or excessively on neighbors' knowledge does not yield good results.

\subsection{Ablation  Study (RQ3)}
\begin{figure*}

\begin{minipage}[t]{0.3\columnwidth}
  \includegraphics[width=\linewidth]{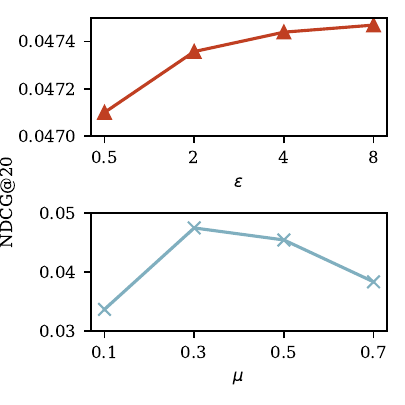}
  \caption{Performance of different hyperparameters on Beijing.}
  \label{fig:hyper}
\end{minipage}\hfill
\begin{minipage}[t]{0.33\columnwidth}
  \includegraphics[width=\linewidth]{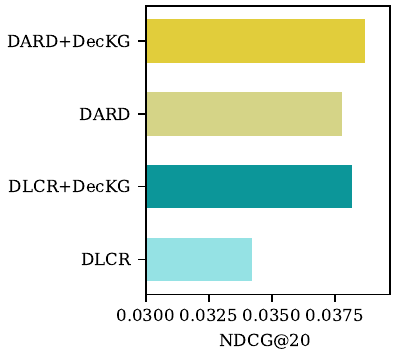}
  \caption{performance of DecKG integrating to other DecPOIs on Beijing.}
  \label{fig:agnostic}
\end{minipage}\hfill
\begin{minipage}[t]{0.3\columnwidth}
  \includegraphics[width=\linewidth]{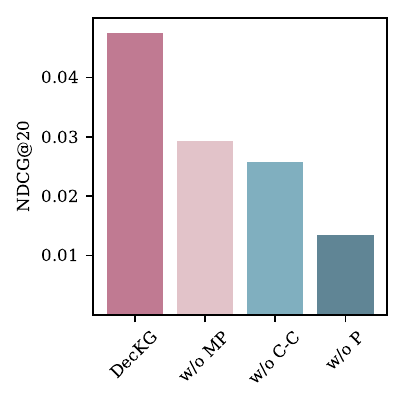}
  \caption{Performance of different DecKG variants on Beijing. }
  \label{fig:ablation}
\end{minipage}
\vspace{-5pt}
\end{figure*}

In this section, we demonstrate the effectiveness of different components of the proposed method, including KG pretraining, client-client communication, and adaptive neighbor sampling. To assess the impact of each component, we conducted ablation experiments where we excluded one component at a time, while preserving the functionality of the others. These experiments were conducted on the Beijing dataset, as illustrated in Fig. \ref{fig:ablation}, with similar trends observed across other datasets.

\textit{w/o MP} refers to the exclusion of the meta-path during knowledge graph partitioning. This approach leads to performance degradation as it only includes the one-hop neighbors of POIs in the sub-knowledge graph, thereby neglecting valuable higher-order information within the KG.

\textit{w/o C-C} discards client-client communication during the device-side refinement stage. In this case, the local model is trained solely on the local check-in data and sub-KGs. A possible explanation for the performance drop is that without sharing user-item interactions, training on local data alone fails to capture the high-order information present in the user-item bipartite graph. In contrast, during client-client communication, this high-order information is implicitly transferred through the exchange of model parameters.

\textit{w/o P} eliminates the knowledge graph pretraining stage on the server. In this setup, after receiving users' desensitized interaction data, the server directly deploys the relevant sub-KGs to users without pretraining. The observed performance degradation highlights the importance of pretraining in achieving optimal results.

\subsection{Model-agnostic Study (RQ4)}

This segment delves into the potential for the proposed DecKG framework to be effectively and seamlessly incorporated with other DecPOIs methodologies, which could facilitate a cohesive interaction among these decentralized frameworks, thereby enhancing the overall efficacy and utility of DecPOIs. The experiment is conducted on Beijing dataset, as illustrated in Figure \ref{fig:agnostic}.

DecKG can be effortlessly incorporated into various decentralized POI approaches. For instance, DLCR transfers model parameters during client-client communication, while DARD uses reference data as a bridge to facilitate communication between neighbors. Both methods can be integrated with the proposed framework, resulting in a noticeable improvement in model performance. This demonstrates the positive effect of incorporating KG, as the high-order information provides valuable auxiliary information to classical DecPOI methods. Furthermore, the results highlight the model-agnostic nature of the proposed framework, confirming that it is a general approach capable of being integrated with both existing and future DecPOI methods.

\section{Conclusion}
This paper presented DecKG, a decentralized collaborative learning framework that enhances POI recommendations by integrating knowledge graph. DecKG addresses the challenges of limited local data and privacy concerns by generating desensitized check-in data and deploying user-specific sub-KGs for local refinement and client-client communication. Our experiments demonstrate that DecKG  improves recommendation performance while ensuring privacy preservation and efficient use of device resources. The framework is adaptable and can be seamlessly integrated with other decentralized POI methods, offering a promising solution for privacy-conscious, high-performance POI recommendations.

\section*{Acknowledgement}
This work is partially supported by the National Key R\&D Program of China under the Grant No. 2023YFE0106300 and 2017YFC0804002, Australian Research Council under the streams of Future Fellowship (Grant No. FT210100624), Discovery Early Career Researcher Award (Grant No. DE230101033), Discovery Project (Grant No. DP240101108), Linkage Project (Grant No. LP230200892), and National Science Foundation of China under Grant No. 62250710682 and 61761136008.

\bibliographystyle{elsarticle-harv} 
\bibliography{main}





\end{document}